\documentclass[conference]{IEEEtran}
\IEEEoverridecommandlockouts
\usepackage{indentfirst}
\usepackage{caption}
\usepackage{subcaption}
\usepackage{cite}
\usepackage{amsmath,amssymb,amsfonts}
\usepackage{algorithmic}
\usepackage{graphicx}
\usepackage{textcomp}
\usepackage{xcolor}
\usepackage{soul}
\usepackage{booktabs}

\def\BibTeX{{\rm B\kern-.05em{\sc i\kern-.025em b}\kern-.08em
  T\kern-.1667em\lower.7ex\hbox{E}\kern-.125emX}}
\begin{document}

\addtolength{\textfloatsep}{-5mm}

\IEEEaftertitletext{\vspace{-1\baselineskip}}

\title{\huge{VECOM: Variation-Resilient Encoding and Offset Compensation Schemes for Reliable ReRAM-Based DNN Accelerator}\vspace{-0.5em}}

\author{\IEEEauthorblockN{Je-Woo Jang\IEEEauthorrefmark{1}, Thai-Hoang Nguyen\IEEEauthorrefmark{2} and Joon-Sung
Yang\IEEEauthorrefmark{1}\IEEEauthorrefmark{3}} \IEEEauthorblockA{\IEEEauthorrefmark{1}School of Electrical and Electronic Engineering and \\
\IEEEauthorrefmark{3}Dept. of Semiconductor Systems Engineering, Yonsei University, Seoul, South Korea  \\ 
\IEEEauthorrefmark{2}Dept. of Electrical and Computer Engineering, Sungkyunkwan University, Suwon, South Korea\\
jeus63@yonsei.ac.kr, th.nguyen@g.skku.edu,
js.yang@yonsei.ac.kr\\[-2.3ex]} 

}

\maketitle

\begin{abstract} 
    Resistive Random-Access Memory (ReRAM)-based Processing In-Memory (PIM) Accelerator has emerged
    as a promising computing architecture for memory-intensive applications, such as Deep Neural
    Networks (DNNs). However, due to its immaturity, ReRAM devices often suffer from various
    reliability issues, which hinder the practicality of the PIM architecture and lead to a severe
    degradation in DNN accuracy. Among various reliability issues, device variation and offset current
    from High Resistance State (HRS) cell have been considered as major problems in a ReRAM-based PIM architecture.
    Due to these problems, the throughput of the ReRAM-based PIM is reduced as fewer wordlines are activated. In
    this paper, we propose VECOM, a novel approach that includes a variation-resilient encoding 
    technique and an offset compensation scheme for a robust ReRAM-based PIM 
    architecture. The first technique (i.e., VECOM encoding) is built based on the analysis
    of the weight pattern distribution of DNN models, along with the insight into the ReRAM's variation property. The second technique, VECOM offset compensation, tolerates offset current in PIM by mapping the conductance of each Multi-level Cell (MLC) level added with a specific offset conductance.
     Experimental results in various DNN models and datasets show that the proposed techniques can increase the throughput of the PIM architecture by up to 9.1 times while saving 50\% of energy consumption without any software overhead. Additionally, VECOM is also found to endure low R-ratio ReRAM cell (up to 7) with a negligible accuracy drop. 

\end{abstract}


\IEEEpeerreviewmaketitle

\section{Introduction}

Processing-In Memory (PIM) architecture, utilizing Resistive Random Access Memory (ReRAM),
offers a promising solution to address the limitations of conventional von Neumann architecture \cite{shafiee_nag2016, 10.1145/3007787.3001140, joshi2020accurate}. The core idea of a
ReRAM-based PIM architecture is to realize ReRAM devices as both storage devices and matrix-multiply-accumulate operation (i.e., MAC) elements. By aligning the cells in a crossbar-like structure,
the MAC operations can be computed in situ with
$\mathcal{O}(1)$ time complexity, making such architecture highly efficient for Deep Neural Network's (DNN) computations.
Despite being a potential candidate, ReRAM-based PIM architecture is hampered by various
reliability issues due to its immature manufacturing process, such as the issue of conductance
variation \cite{lin_wen2021} and non-ideal current \cite{park_lee2020}. Such issues can impede the
performance of the ReRAM-based PIM, undermine the DNN accuracy and pose numerous
challenges for the deployment of high reliability-constrained DNN applications.

One of the primary concerns in ReRAM-based systems is conductance variation, where the actual analog conductance value written to a ReRAM cell deviates from the desired data. This issue is particularly notable in the PIM architecture where errors can be accumulated along the bitlines of the crossbar array, leading to incorrect output current. To diminish the effect of accumulated variation errors, existing ReRAM-based PIM architectures often limit the maximum number of activated wordlines (MAW) \cite{park_lee2020}. However, reducing the MAW also curtails the potential for parallelism in a PIM architecture, consequently limiting throughput.
In an attempt to address ReRAM's variation, several works have proposed solutions for ReRAM-based PIM accelerators \cite{lin_wen2021, chen2017accelerator, liu2019fault, lastras2021ratio, park_lee2020}. Notably, software-based techniques \cite{chen2017accelerator, lin_wen2021, liu2019fault} have aimed to lessen the impact of ReRAM's variation by retraining or fine-tuning the DNNs to adapt to the non-ideal distribution of ReRAM resistance. However, these methods can be cost-prohibitive due to the required re-training processes, which is problematic when the DNN is further scaled or when DNN is already deployed to edge devices. 
Alternatively, other approaches introduce redundant hardware to compensate for variation \cite{park_lee2020, chen2019cmos}, but these techniques often result in significant hardware overhead, making them unsuitable for resource-constrained devices.

In addition to the variation, the offset current generated when an input voltage is applied to High Resistance State (HRS) cells can pose a significant challenge in parallelizing MAC operations of ReRAM-based accelerators. In an ideal scenario, the output current of HRS cells should be zero during MAC operations. However, non-ideal ReRAM devices exhibit offset current leakage from HRS cells when an input voltage is applied. This issue becomes more pronounced when utilizing low R-ratio ReRAM cells, which have a low ratio of resistance values between the on and off states. Several works have been proposed to ensure reliable operations with a low R-ratio ReRAM device \cite{8715178, 9756843, park_lee2020,chen201865nm, chen2019cmos}. 
Among them, \cite{chen2019cmos} proposes a distance-racing readout scheme to mitigate the adverse effect from the offset current by measuring the distance of accumulated current on a bitline from two reference sensing currents. However, the number of reference currents is proportional to the number of activated wordlines. Hence, the MAC operation in a typical crossbar size of 128x128 \cite{shafiee_nag2016} is not feasible due to the high area and power overhead of the peripheral circuit. The work in \cite{park_lee2020} subtracts the offset current by adding an extra column of HRS cells. While proven effective, such a method is limited to Single Level Cells (SLC) ReRAM and difficult to apply to higher ReRAM cell-precision such as Multi-level Cell (MLC).

In this paper, we introduce VECOM, which includes a variation-resilient encoding technique and an offset current compensation scheme for a robust ReRAM-based DNN accelerator. 
The contributions of this paper can be summarized as follows:

\begin{itemize}
    \item We analyze the characteristics of weight patterns in various DNN models and their impact on the performance of PIM-based DNN systems.
    \item We propose VECOM encoding, a two-step encoding technique, to enhance the robustness of DNNs against ReRAM variations. The proposed encoding scheme is found to be effective, lightweight, and significantly improves the throughput of the PIM architecture.
    \item We introduce a programming scheme of VECOM which can be utilized for low R-Ratio and various cell-precision (number of bit-per cell) of ReRAM device. This approach effectively mitigates offset current from the HRS cells without adding the burden of latency and power to the conventional programming techniques.
    \item VECOM is evaluated using various DNN models and datasets. The results demonstrate that VECOM outperforms the baseline PIM design and other existing techniques in terms of enhancing DNN robustness and increasing wordline-parallelism of non-ideal ReRAM-based PIMs.

\end{itemize}

The rest of the paper is organized as follows. The backgrounds and related works for PIM accelerator
using MLC ReRAM, as well as the problem of variation and offset current in the ReRAM-based PIM are presented in Sec. \ref{sec:background}. Sec. \ref{sec:proposed-techniques} presents the analysis of pattern distribution in DNNs and its impact on DNN accuracy. The proposed VECOM encoding technique and conductance offset mapping programming are also
introduced in this section. Evaluations of the proposed method and a detailed discussion of the results are given in Sec. \ref{sec:evaluation}. Sec. \ref{sec:conclusion} concludes the paper.

\section{Backgrounds and Related Works}\label{sec:background} 

\subsection{ReRAM-based DNN Accelerator}\label{subsec:reram-pim}

\begin{figure}[t]
\centerline{\includegraphics[width = 1.1\linewidth]{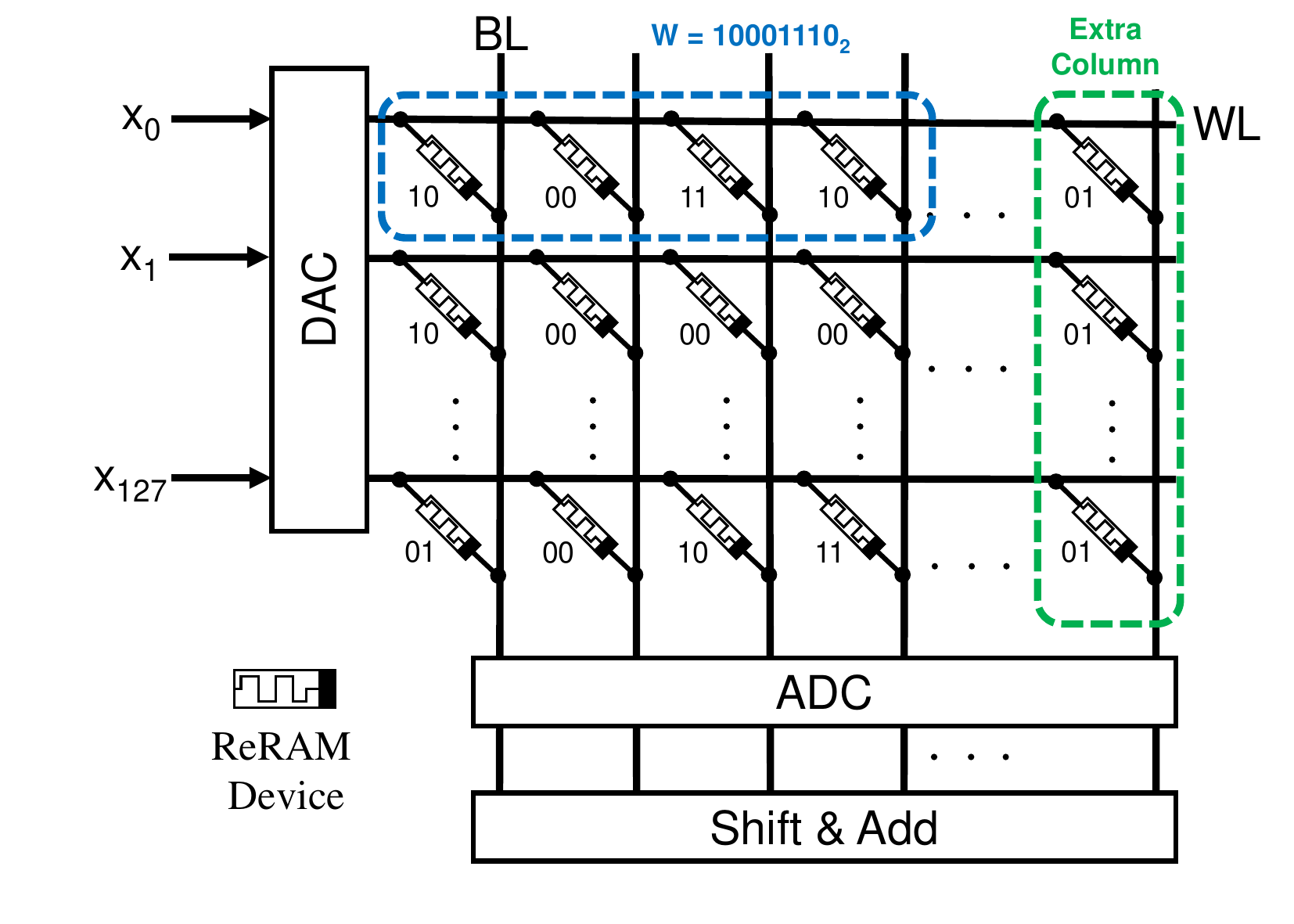}}
\vspace{-0mm}
\caption{An example of a typical ReRAM-based crossbar array. All signed 8-bit weights are
     represented using the unsigned format with the additional bias.}
\label{fig:reram-accelerator}

\end{figure}

Deep Neural Networks (DNNs) have demonstrated exceptional performance and efficiency across numerous machine learning applications. As DNNs continue to evolve, there is an escalating demand for more robust and scalable hardware architectures. To meet this need, ReRAM-based PIM, a high-throughput, energy-efficient hardware accelerator for DNNs, has emerged as a viable solution for future DNN computations \cite{shafiee_nag2016, shin2021fault}. Fig. \ref{fig:reram-accelerator} depicts a typical array-level configuration of a ReRAM-based DNN accelerator. This PIM architecture includes multiple ReRAM cells organized in a crossbar configuration, along with peripheral circuits such as digital-to-analog converters (DACs), analog-to-digital converters (ADCs), and shift-and-add (S\&A) components. Each ReRAM cell is connected to a wordline (WL) on one side and a bitline (BL) on the other. When an input voltage is applied to each wordline, the accumulated current at the bitline's end varies depending on the conductance value of each ReRAM cell. In this discussion, it's assumed that a single ReRAM cell can store 2-bit data (MLC), a configuration prevalent in most ReRAM-based designs \cite{shafiee_nag2016, 10.1145/3307650.3322271, shin2021fault}.

In a ReRAM-based PIM architecture, an 8-bit quantized weight of the DNN model can be represented by 4 MLC ReRAM cells. Since PIM architecture cannot represent the
sign of the DNN's weights, existing PIM architecture, e.g., ISAAC \cite{shafiee_nag2016}, uses biased representation for weights and adds an extra unit column to count the number of biases in input and then subtract them from the output digital value to compute signed arithmetic operation. For example, consider the case of a weight located in the first row of the crossbar array illustrated in Fig. \ref{fig:reram-accelerator}. The biased weight W, located inside the blue bounding box, is expressed with $10001110_2$ in binary which is $142_{10} = 128_{10} + 14_{10}$ in decimal. Since the 8-bit fixed-point quantized weight is in the range of -128 to +127, when a bias value of +128 is added, all weights mapped into the crossbar array have values ranging from 0 to +255, which can be represented by a positive ReRAM conductance value. Since biases are added to the weights, the results of MAC operation from the crossbar array are also added with the bias values. These bias values must be subtracted in order to obtain the signed arithmetic values. The input vector ($x_{0 \text{--} 127}$) applied to the crossbar array is also applied to the extra column in Fig. \ref{fig:reram-accelerator}. All cells in the extra column are mapped to 01, and the accumulated current at the end of the extra column is converted to a digital value by an ADC to count the number of 1s in the input. After multiplying the number of 1s by the bias and subtracting it from the MAC operation result, the identical MAC operation result using the weight without bias can be obtained.\\

\begin{figure}[t]
\centerline{\includegraphics[width = 1.1\linewidth]{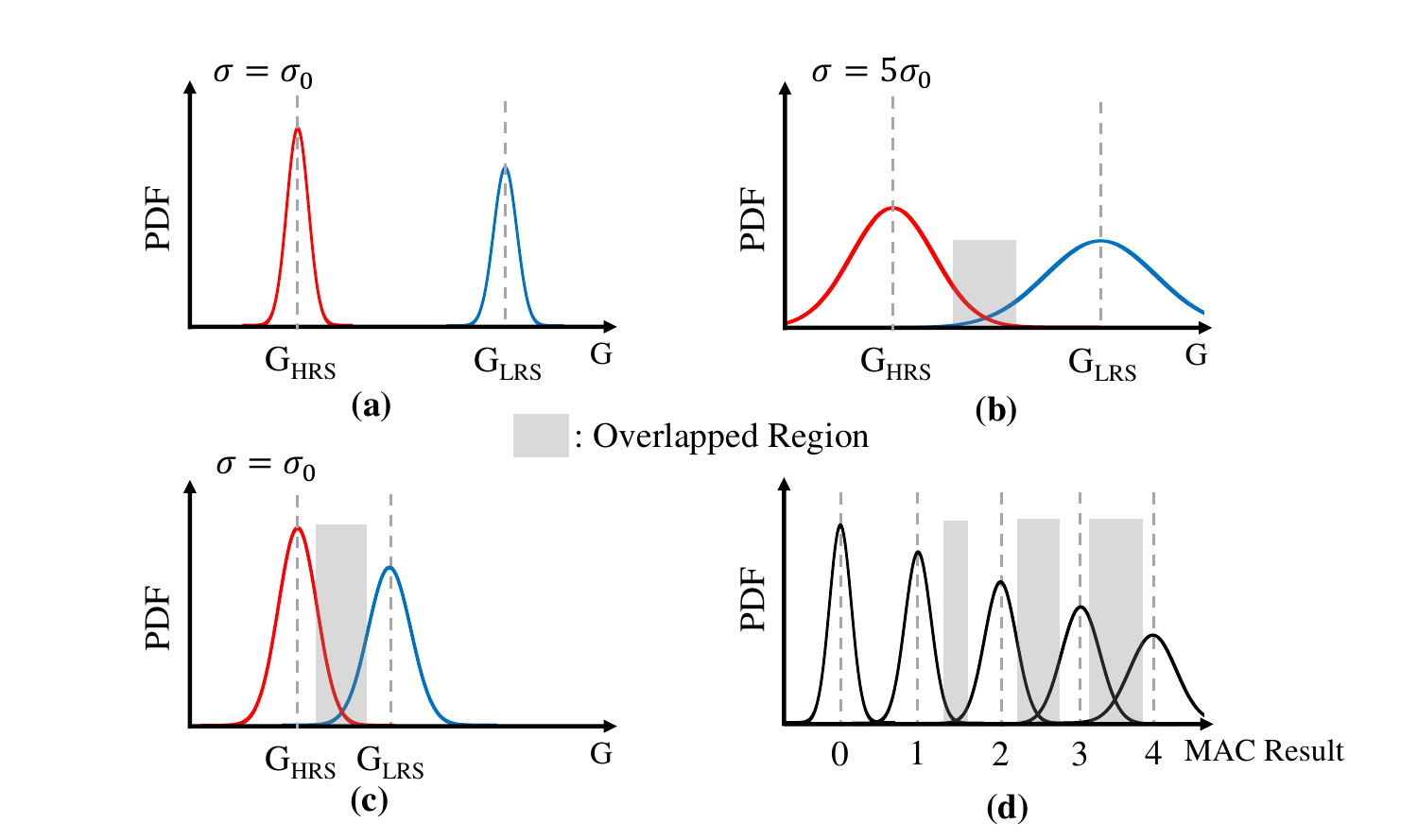}}
\vspace{-2mm}
\caption{Non-ideal properties of ReRAM cell limiting parallel activation of wordlines compared to the ideal case. (a) High R-ratio with low variation. (b) With large variation. (c) With a low R-ratio. (d) MAC current distribution. }
\label{fig:reram-property}

\end{figure}

\subsection{Non-Ideal Properties in ReRAM-based PIM Architecture}

Although being effective in processing the MAC operations, ReRAM-based PIM 
architecture suffers from various reliability issues caused by the immaturity of the manufacturing
process and its analog characteristics. Fig. \ref{fig:reram-property} depicts the non-ideal properties that occurred in the operations of ReRAM. Fig. \ref{fig:reram-property} (a) shows the conductance distribution using the ideal ReRAM which has a large R-ratio and a low cell variation. Fig. \ref{fig:reram-property} (b) and (c) respectively show the ReRAM cell with large variation and low R-ratio. As shown in the figures, cell variation and low R-ratio cause an overlap of the conductance distributions between neighboring levels. This can lead to the overlap in the MAC result current distributions, as illustrated in Fig. \ref{fig:reram-property} (d). As a result, the overlapped region causes the ADC's logic to be ambiguous when distinguishing between discrete digital values, which is detrimental to the system.

Previous works \cite{park_lee2020, chen2019cmos}
often limit the maximum number of activated wordlines (MAW) during the operation of PIM, since turning on a large number of
wordlines can cause wrong outcomes at the end of each column. Such an approach reduces the PIM's throughput in each cycle and thus degrades the performance of the system. 
Therefore, it is necessary to develop an encoding technique to improve the throughput of a ReRAM-based PIM architecture while still keeping the impact of ReRAM variation minimal.

\begin{figure*}[t]
\centerline{\includegraphics[width = 0.8\linewidth]{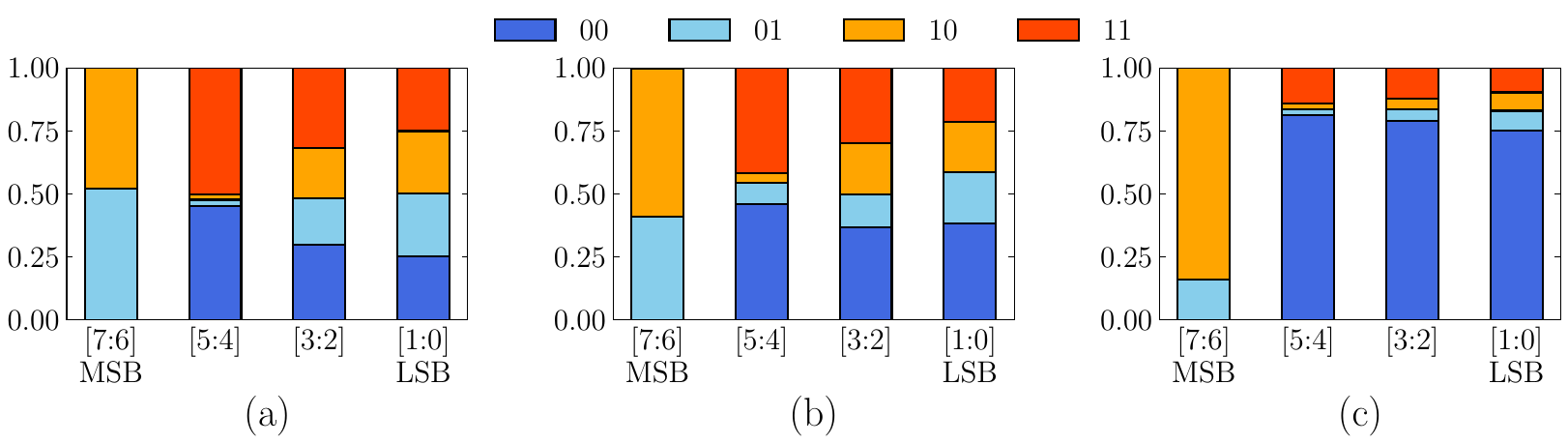}}
\caption{MLC level distribution in various DNN models on CIFAR-10. (a) ResNet-18. (b) VGG-16, (c) Inception-V3.}
\label{fig:level_distribution}
\end{figure*}

\begin{figure}[t]
    \centering
    \centerline{\includegraphics[width = .8\linewidth]{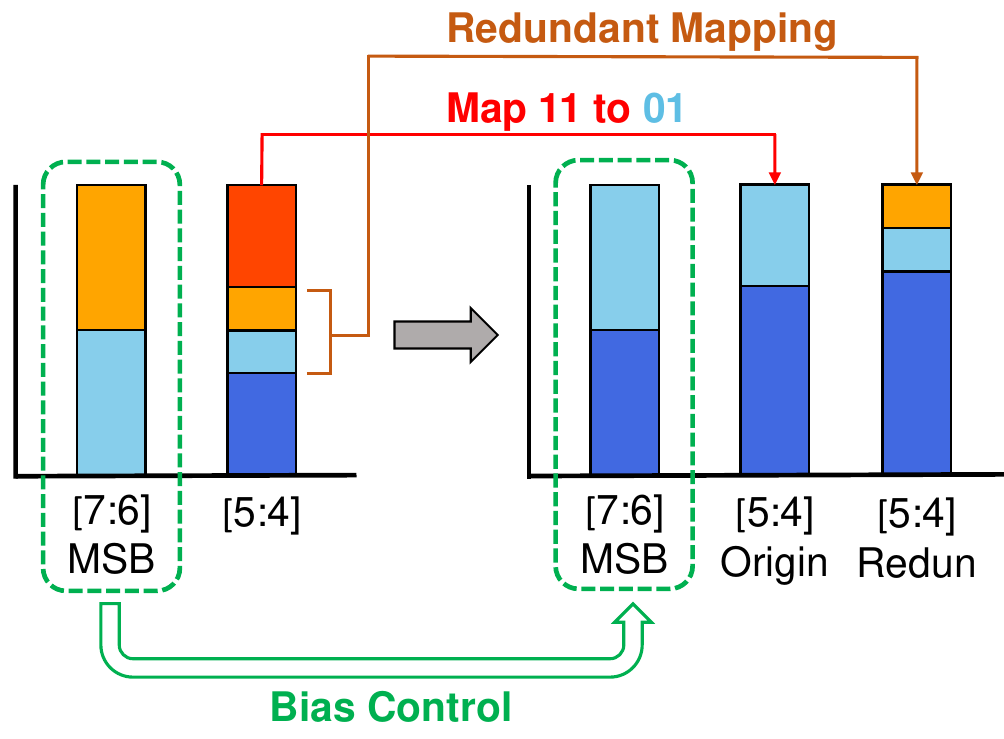}}
    \vspace*{-1mm}
    \caption{Overview of the proposed variation-resilient encoding technique} 
    \label{fig:encoding_flow} 
\end{figure}

\subsection{Related Works} Various techniques have been proposed to resolve the problem of variation
in ReRAM-based PIM architecture \cite{lin_wen2021, chen2017accelerator, liu2019fault,
lastras2021ratio, park_lee2020}. Specifically, the works in \cite{chen2017accelerator, lin_wen2021,
liu2019fault} utilize the self-healing capability of DNN to compensates the variation caused by ReRAM
cells. In other words, re-training the network with the consideration of ReRAM variation is required
for such techniques. Although being effective, these approaches are not feasible when the DNN has
already been deployed and the training data is not available. Other hardware-based approaches have
tried to reduce the impact of variation by using redundant hardware designs \cite{chen2019cmos,
park_lee2020}. \cite{park_lee2020} proposes a method that counts the number of activated wordlines
in the digital domain and then compensate the corresponding variation using ADCs. In this manner,
\cite{chen2019cmos} requires a separate ReRAM array with the same size as the original array, to
store the ADC reference current information. This imposes a considerable overhead which would not be
suitable in resource-constrained environments. There are also studies that use Error Correcting Code (ECC) to increase the variation-tolerability of the ReRAM-based PIM accelerators. \cite{successive_correction, ancode, ldpc} have used ECC such as AN code, Low Density Parity Check (LDPC) or successive error correction to lessen variation in the ReRAM-based PIM. Although proven effective, such techniques require a large power and latency overhead to encode and decode data, as well as to correct MAC operation values. Another work \cite{gounary} has utilized unary encoding to reduce the overhead of retraining. \cite{gounary} proposes priority unary mapping based on the fact that the conductance value $G$ of a ReRAM cell under variation is often modeled by the following:
\begin{equation}
    \label{conductance_eq}
    \begin{split}
G & = G_0 \cdot e^{\theta} \qquad \theta \sim N(0, \sigma^2) \\
    \end{split}
\end{equation}

Where $G_0$ is a conductance value of the ReRAM cell without any variation and the conductance value $G$ follows the log-normal distribution with zero mean and $\sigma$ variation. 
Even though this method can mitigate the influence of variation on DNN accuracy, it nonetheless results in a 250\% area overhead compared to standard mapping. This makes it impractical for use with resource-limited edge devices.

Apart from the variation issue, numerous studies have suggested solutions for the offset current problem through the use of devices with low R-ratios\cite{9756843, park_lee2020, 8715178}. Input-Aware Current Compensation (IAC) technique in \cite{park_lee2020}
adds one column of HRS cells to the crossbar array to compensate for the offset current of the existing ReRAM cells. However, in this manner, the result of PIM could be inaccurate if R-ratio is not sufficiently larger than guaranteeing the negligible loss by the compensation. Moreover, it is only applicable to Single-Level Cell (SLC) configurations. Consequently, it becomes essential to design methods capable of reliably conducting MAC operations in conditions of significant variation and low R-ratios, without necessitating expensive re-training processes or inducing substantial hardware overhead.

\section{Proposed Method}\label{sec:proposed-techniques}

\subsection{Proposed Variation-Resilient Encoding Technique}\label{subsec:proposed_encoding} 

To enhance the robustness of DNNs under the impact of cell variation, we first analyze the pattern distribution of DNN's parameters when using MLC ReRAM cells. Fig. \ref{fig:level_distribution} depicts the MLC level distributions of 8-bit fixed-point quantized weights in ResNet18, VGG16, and InceptionV3. As mentioned in Sec. \ref{sec:background}, a DNN weight can be represented using four MLC ReRAM cells from the MSB of [7:6] to the LSB of [1:0].

As shown in Fig. \ref{fig:level_distribution}, while various MLC level distribution patterns are found from [5:0] bit positions, there are mostly 10 and 01 patterns existing in the [7:6] MSB position.
With the 8-bit fixed point quantization, weights possessing MSB of either 01 or 10 fall within the range of -64 to +64. This interval accommodates the majority of weights within the quantization range.
On the other hand, the pattern 00 and 11 belong to outlier values \cite{9774543}, which explains why these patterns only take up a small amount in the MSB position. According to Eq. \ref{conductance_eq}, the conductance value representing 10 is significantly large, potentially leading to a substantial overlap region in the MAC results distribution \cite{gounary}.
This could subsequently induce inaccurate values in the ADC output for the MSB column, severely impairing DNN accuracy. The error in the MSB position is often found to be more significant to DNN's accuracy compared to other bit positions in the DNN's parameter \cite{8806855, nguyen2022dynapat}. Therefore, by minimizing the overlapped region of output current distribution, caused by the conductance variation of the MSB, we can enhance the model's robustness against variation.

\begin{table}[]
  \centering
  \captionof{table}{Ratio of clipped weights when utilizing VECOM across different DNN models.\label{Tab:Tcr}}
\begin{tabular}{lllll}
\cmidrule{1-4}
\multicolumn{1}{|c|}{{Model}}               & \multicolumn{1}{c|}{ResNet-18} & \multicolumn{1}{c|}{VGG-16}  & \multicolumn{1}{c|}{Inception-V3} &  \\ \cmidrule{1-4}
\multicolumn{1}{|c|}{Clipped Weight Ratio (\%)} & \multicolumn{1}{c|}{0.00059}   & \multicolumn{1}{c|}{0.00018} & \multicolumn{1}{c|}{0.002}        &  \\ \cmidrule{1-4}
\end{tabular}
\vspace{-1mm}
\end{table}

As mentioned in Sec \ref{sec:background}, a bias value of 128 is added to the DNN's weights to represent 8-bit signed values. However, based on the observation in Fig. \ref{fig:level_distribution}, most patterns in the MSB position are either 01 or 10, which makes almost all of the mapped weights fall into the $\pm$64 range. Therefore, in this case, it is not essential to add 128 as bias. Drawing from this insight, we propose a method to convert pattern 01 and 10 to pattern 00 and 01 by adjusting the bias value during the weight mapping process. The VECOM encoding scheme, depicted in Fig. \ref{fig:encoding_flow}, involves two main processes: bias control and redundant mapping. 
To transform the MSBs into a pattern of 00 and 01, the bias is adjusted downwards to 64, a process referred to as bias control within the proposed VECOM encoding. This modification remains weight values below 0 which cannot be mapped to the crossbar array. Therefore, these weights are clipped to zero and mapped to the PIM architecture. This approach is akin to conventional weight clipping \cite{joardar2021learning}, but with a bias of 64, only negative values are clipped. Our proposed method mitigates the typical accuracy loss associated with conventional weight clipping. 
This becomes evident when examining the portions of weights that are clipped by the proposed method. As demonstrated in Table \ref{Tab:Tcr}, only an insignificant 0.002\% of the total weight portion is affected by our method, thus it does not induce any notable degradation in the DNN accuracy.

Following bias control, a redundant mapping process is performed to improve the robustness of the [5:4] position. In a similar fashion to the previous step, to map higher level cells to lower level cells, we map a minor portion of the 01 and 10 patterns to a redundant array and fill this with 00, as illustrated in Fig. \ref{fig:encoding_flow}. Subsequently, we substitute the mapped 00 and 11 patterns with 00 and 01 patterns. Through this process, the patterns in the [5:4] position are distributed into two arrays. The original [5:4] array, labeled 'Origin' in the figure, consists of 00 and 01, whereas the redundant [5:4] array (labeled 'Redun') contains patterns of 00, 01, and 10. When applying the proposed method, the MAC results of the original [5:4] array, which are passed through the ADC in the S\&A stage, are multiplied by 3 to obtain the identical result as the existing mapping. Therefore, the proposed method enables reliable operation by converting the [7:4] MSB, which heavily influences the accuracy, into patterns exhibiting less overlap in the MAC result current distribution. This leads to improved variation tolerance and enables a high level of wordline parallelism. Since the proposed method operates at the encoding stage without necessitating additional DNN model retraining, there is no time and energy consumption caused by the retraining process.

\begin{figure}[t]
    \centering
    \vspace{-5mm}
    \centerline{\includegraphics[width = 1.0\linewidth]{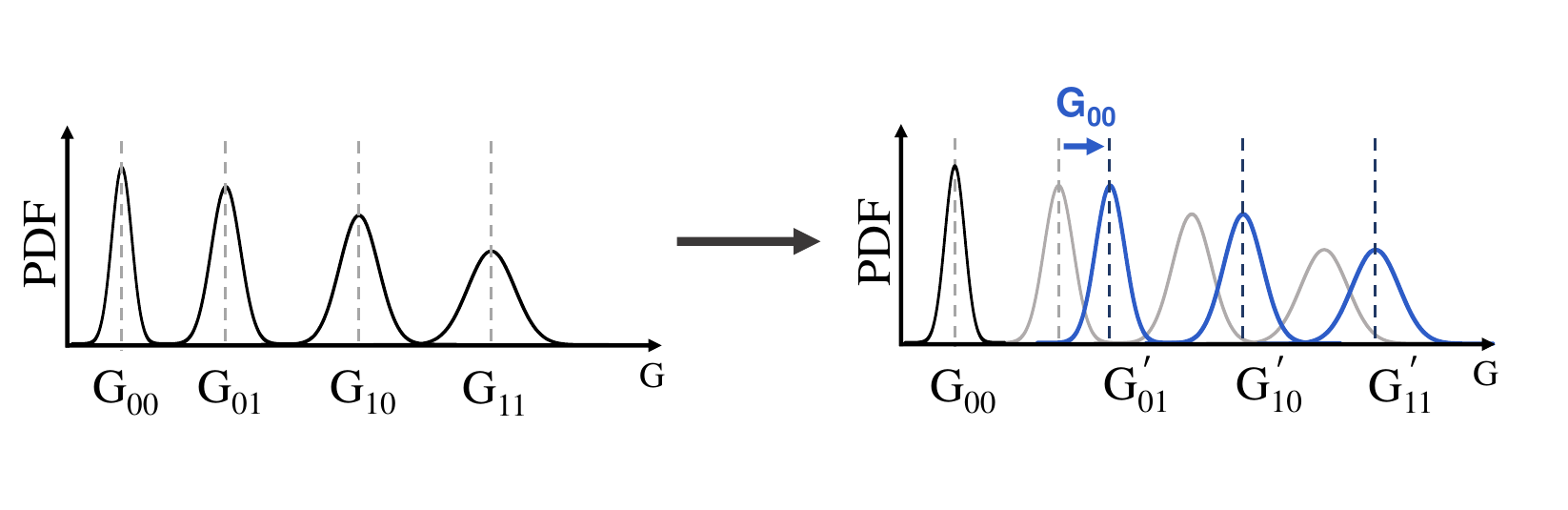}}
    \vspace{-3mm}
    \caption{Overview of the proposed conductance offset mapping for VECOM} 
    \label{fig:COM_flow} 
\end{figure}

\subsection{Proposed Offset Compensation Scheme}\label{subsec:proposed-com}

IAC \cite{park_lee2020} addresses the issue of offset current in ReRAM-based PIM architecture by adding an extra HRS column. Assume the number of activated wordlines in the crossbar array is $N$, with $N_1$ being the number of activated LRS cells among N cells, and $N_0$ cells are in HRS. Under these conditions, the current $I_{BL}$ that accumulates in each bitline can be calculated as follows:
\begin{equation}
    \label{Eq1}
    \begin{split}
I_{BL} & = N_{1} \cdot I_{1} + N_{0} \cdot I_{0}  \\
& = N_{1} \cdot (I_{1} - I_{0}) + N \cdot I_{0} \\
\therefore \quad I_{BL} &- N \cdot I_{0} \approx N_{1} \cdot I_{1} 
    \end{split}
\end{equation}
\indent Here, the bitline current value of the extra column for IAC matches $N \cdot I_{0}$. Hence, the MAC value in each column can be approximated by subtracting the output current of the extra HRS column. According to Eq. \ref{Eq1}, in the case of SLC, the current of the LRS cell is larger than the current of the HRS cell, thus IAC can function stably at a relatively low R-ratio.
However, this is not the case when applying to MLC ReRAM cell. When employing MLC, if the numbers of activated cells correspond to each level from $N_{11}$ to $N_{00}$, the current flowing on the bitline $I_{BL}$ is calculated as follows:  
\begin{equation}
    \label{Eq2}
    \begin{split}
I_{BL} & = N_{11} \cdot I_{11} + N_{10} \cdot I_{10} + N_{01} \cdot I_{01} + N_{00} \cdot I_{00} \\
& = \sum_{i=01}^{11} N_{i} \cdot (I_{i} - I_{00}) + N \cdot I_{00}\\
& \not\approx \sum_{i=01}^{11} N_{i} \cdot I_{i}  + N \cdot I_{00}\\
    \end{split}
\end{equation}
\indent In this case, accurately approximating the currents of the two intermediate states poses a challenge. As the currents from intermediate state cells of 01 and 10 diminish due to offset subtraction, the unexpected reduction in the MAC results undermines the DNN accuracy. 
To address this, we propose a conductance offset mapping of VECOM which ensures effective current compensation in MLC ReRAM devices.
The proposed method adjusts the target conductance during the programming process by adding the conductance value of the lowest MLC level to the other three levels as an offset conductance, i.e., $G' = G + G_{00}$, as shown in Fig. \ref{fig:COM_flow}. The output current after integrating this offset conductance, $I_{BL,COM}$, can be calculated as follows:

\begin{equation}
    \label{Eq3}
    \begin{split}
    I_{BL,COM} &= \sum_{i=01}^{11} N_{i} \cdot (I_{i} + I_{00}) + N_{00} \cdot I_{00} \\
  &= \sum_{i=01}^{11} N_{i} \cdot I_{i} + N \cdot I_{00}\\
  \therefore \quad I_{BL,COM} & - N \cdot I_{00} = \sum_{i=01}^{11} N_{i} \cdot I_{i} 
    \end{split}
\end{equation}

By adjusting the target conductance, we can accurately calculate the offset-compensated bitline current without any approximation. The target conductance value can be easily adjusted during the programming process using the conventional write-and-verify programming scheme \cite{ryu2018optimized}. Notably, the proposed offset compensation scheme operates independently of the cell precision, which means it can be applied to MLC or even higher cell-precision options such as Triple-Level Cell (TLC), Quad-Level Cell (QLC), and so forth. Furthermore, since it uses the current subtraction used in IAC, the proposed method does not introduce any performance degradation at runtime.

\begin{figure}[t]
\centerline{\includegraphics[width = 1\linewidth]{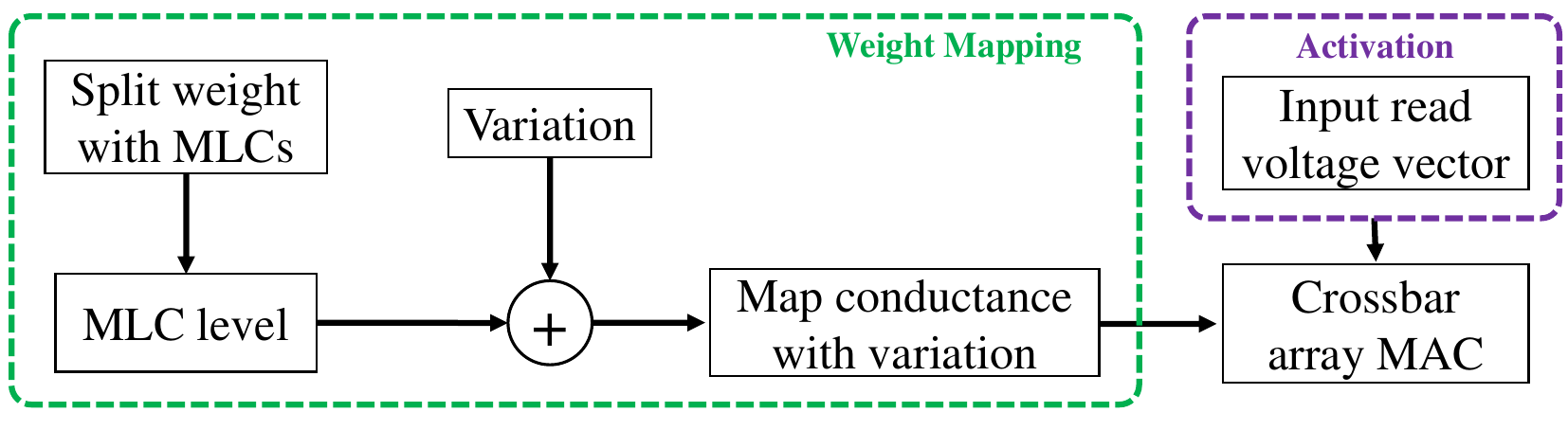}}

\caption{Simulation flow for MLC ReRAM crossbar array computation.}

\label{fig:evaluation_flow}
\end{figure}

\section{Evaluation}\label{sec:evaluation}

\subsection{Experimental Setups} 

To examine the performance and robustness of the proposed VECOM technique, we conduct simulations based on the PyTorch framework. The simulations focus on a ReRAM-based DNN accelerator with a hardware configuration based on ISAAC architecture\cite{shafiee_nag2016}. In order to assess variation resilience, we quantize the activations with 8 bits and the weights with 4 bits. We also use 8-bit weight quantization to determine the maximum number of activated wordlines at a specific variation value.

During the simulations, each bit of the input activation is sequentially fed into the computation unit through a 1-bit DAC. Each computation unit consists of a 128x128 crossbar array of ReRAM cells, with each cell having a precision of 2 bits per cell. The accumulated analog current along the bitline was then converted into a digital value through an ADC.
These evaluations aim to assess the performance and robustness of VECOM across a variety of models, including VGG-16, ResNet-18, and Inception-V3, on various datasets such as CIFAR-10, CIFAR-100, and ImageNet.

We carry out a Monte-Carlo based simulation as shown in Fig. \ref{fig:evaluation_flow}.
The first step involves decomposing the quantized weights into multiple MLCs, based on the position of each 2-bit value. To incorporate variation, a log-normal distribution is utilized (as shown in Eq. \ref{conductance_eq}), where the variation parameter is consistent across all levels of conductance. This distribution is then sampled to generate variations in the conductance levels. The quantized activation vector is sequentially fed into the crossbar weight array, facilitating analog matrix-vector multiplication.

It is worth noting that in the VECOM structure, the MLCs comprising a single weight are not mapped to the same array; instead, bits with the same significance are mapped to their corresponding arrays. For example, the MSBs are mapped to one array, while the LSBs are mapped to another array. This approach does not impact the results or performance of the PIM architecture. Additionally, an extra column is added per array to perform current subtraction via conductance offset mapping.
This method only requires one additional column per array, resulting in minor hardware overhead for offset compensation.

\subsection{Results and Discussion}

\begin{figure}[t]
\vspace{-2mm}
\centerline{\includegraphics[width = 0.9\linewidth]{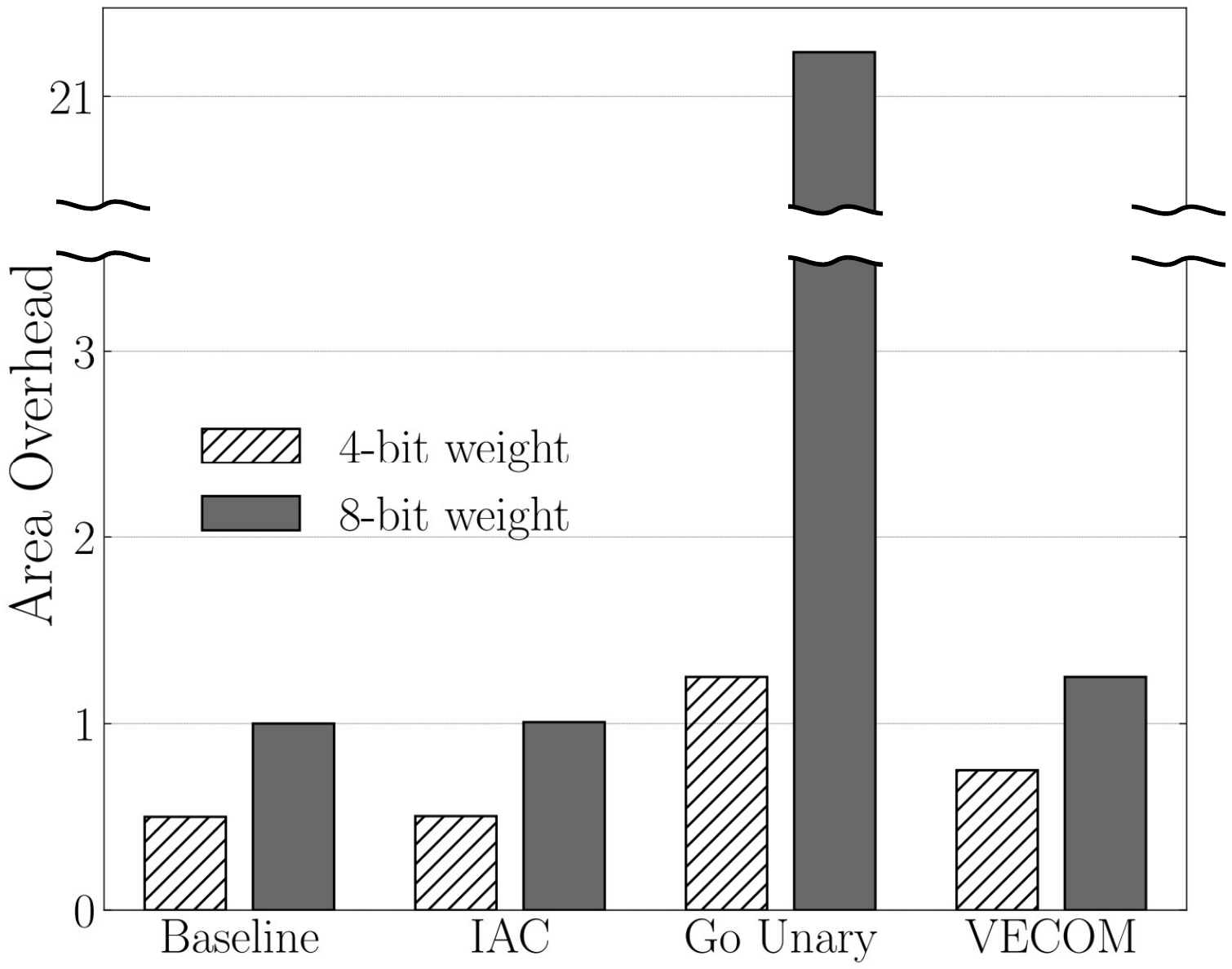}}
\caption{Area Overhead Comparison of VECOM with baseline, IAC and Go Unary.}

\label{fig:compare_area}

\end{figure}

\subsubsection{Hardware Overhead}\hfill

Fig.\ref{fig:compare_area} shows a comparison of the crossbar area overhead introduced by VECOM, the baseline method, and other approaches \cite{gounary, park_lee2020}. 
Specifically, we compare the area overhead resulting from the additional column in the IAC technique \cite{park_lee2020} and the unary encoding used in Go Unary \cite{gounary}, considering both 4-bit and 8-bit weight quantization. The baseline refers to the crossbar area of the conventional mapping scheme, employing an 8-bit weight quantization using MLC. As depicted in the figure, the area overhead of VECOM is approximately 25\% higher than the baseline when using 8-bit quantization, mainly due to the redundant mapping of VECOM.

For the IAC technique, the area overhead is around 1\% because it requires the addition of one column per array. On the other hand, when using Unary encoding with 4-bit quantization, the area overhead is approximately 25\% compared to the baseline. Moreover, with 8-bit quantization, the number of MLCs needed to cover a wider range of weight values increases exponentially, resulting in a nearly 20-fold increase in the area overhead compared to the baseline. While VECOM results in a higher area overhead compared to the baseline, it offers benefits such as increased robustness and enhanced wordline parallelism, contributing to better overall performance and DNN accuracy.

 \begin{figure}[t]

\centerline{\includegraphics[width = \linewidth]{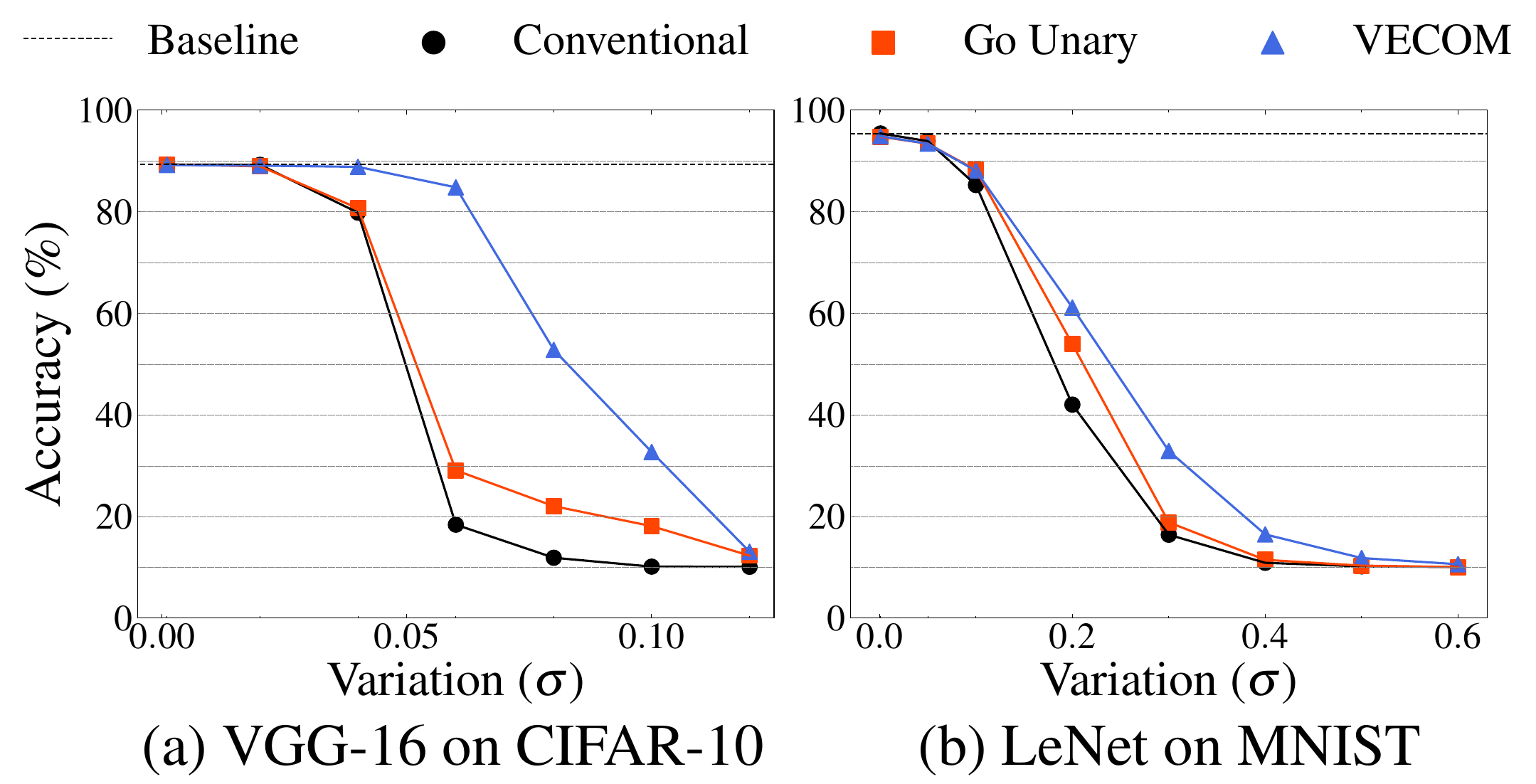}}

\caption{Variation tolerance of VECOM compared to conventional mapping and Go Unary}

\label{fig:compare_unary}

\end{figure}
\subsubsection{Effectiveness on Tolerating Variation and Paralleling Multiplication}\label{sec:variation_tolerance}\hfill

In this section, we conduct a comparison between Unary encoding and the conventional mapping with VECOM to evaluate their tolerance to variation when all 128 wordlines in a crossbar array are activated. To ensure a fair comparison with Go Unary, we apply 4-bit quantization to LeNet trained on the MNIST dataset and VGG-16 trained on the CIFAR-10 dataset. 

\begin{figure*}[t]
\vspace{-6mm}
\centerline{\includegraphics[width = .7\linewidth]{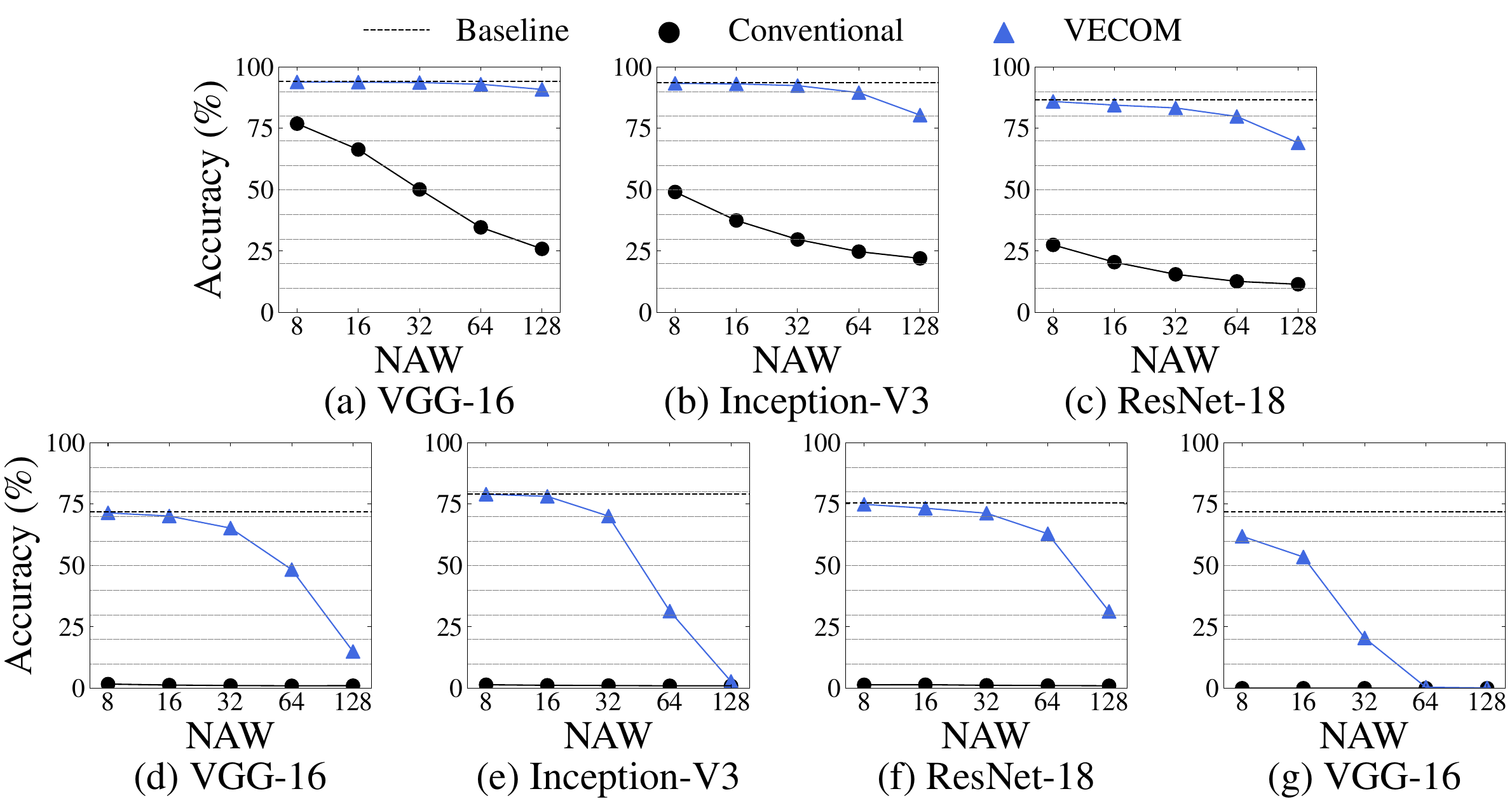}}
\caption{Model accuracy based on the NAW of conventional mapping and VECOM in various datasets. (a) - (c) on CIFAR-10. (d) - (f) on CIFAR-100 and (g) on ImageNet.}
\label{fig:naw_test}

\end{figure*}

The results, depicted in Fig. \ref{fig:compare_unary}, reveal that VECOM exhibits a smaller accuracy degradation compared to both the conventional mapping and the priority mapping proposed by Go Unary under the same degree of variation. VECOM promotes more reliable MAC operations while incurring less area overhead than unary encoding. Notably, in the case of VGG-16, significant accuracy degradation occurs starting from a variation of 0.08.
To further analyze the impact of the number of activated wordlines (NAW) at variation = 0.08, experiments are run.
Fig. \ref{fig:naw_test} presents the model accuracy with varying NAW for ResNet-18, VGG-16, and Inception-V3 on CIFAR-10, CIFAR-100, and ImageNet datasets. To achieve higher accuracy, we utilize 8-bit weight quantization instead of 4-bit quantization. The results show that VECOM consistently outperforms the conventional mapping in terms of accuracy for all datasets and DNN models. For example, on the CIFAR-10 dataset, the accuracy difference reaches up to 58.4\% for NAW = 8 and 64.9\% for NAW = 128. On the CIFAR-100 dataset, the accuracy difference is up to 77.6\% for NAW = 8 and 31.2\% for NAW = 128. Lastly, on the ImageNet dataset, we observe an accuracy improvement of 62\% for NAW = 8. Overall, VECOM allows for a significantly higher number of NAWs while maintaining the same accuracy, thereby improving the parallelism of the multiplication of ReRAM-based DNN accelerators.

\subsubsection{Scalability on R-Ratio and Bit-per-Cell}\label{sec:lowratio}\hfill\\
To examine the stability of VECOM at low R-Ratios with offset current compensation, we compare it with conventional MAC operations and IAC. The R-ratio spans from an ideal value of 1000 \cite{li2015variation} to a low value of 7. For the comparison, we evaluate the accuracy of the ResNet-18 model using Single-Level Cell (SLC), where IAC can take advantage of its current compensation. We also evaluate the accuracy of the VGG-16 model when using MLC. In these evaluations, we set the variation to 0.04, as used in IAC \cite{park_lee2020}, and activate 128 wordlines for each MAC operation.

\begin{figure}[t]
\vspace{-5mm}
\centerline{\includegraphics[width = 0.9\linewidth]{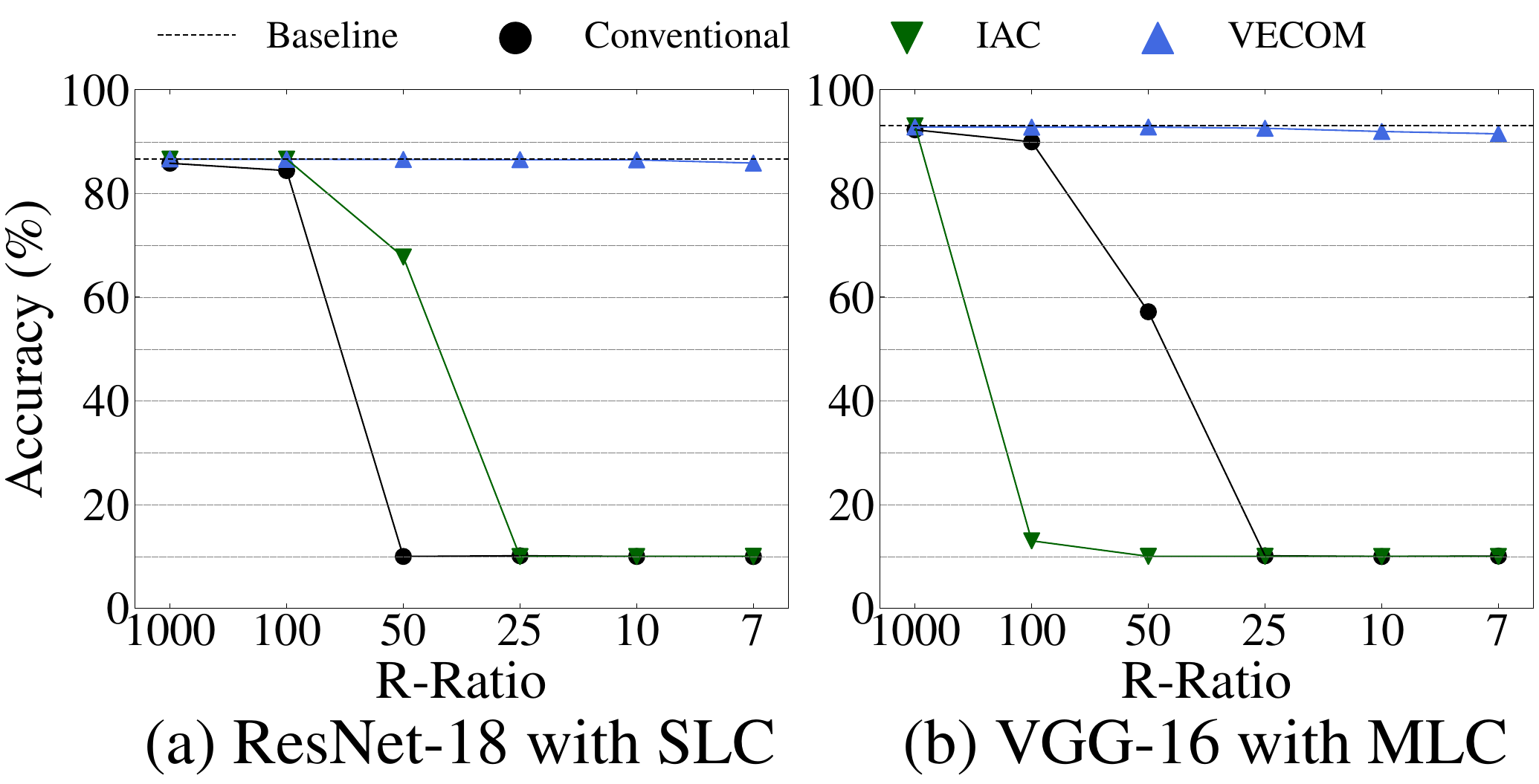}}

\caption{R-Ratio endurance comparison with IAC. (a) ResNet-18 with SLC (b) VGG-16 with MLC on CIFAR-10.}
\label{fig:comp_iac}

\end{figure}

The results, as shown in Fig. \ref{fig:comp_iac}, demonstrate that IAC exhibits a relatively less accuracy degradation at lower R-ratios compared to conventional MAC operations using SLC. On the other hand, VECOM shows almost no degradation even when the R-ratio drops to 7. This verifies the robustness of VECOM in reliably carrying out the MAC operations, even at low R-ratios. Additionally, for VGG-16 with MLC, IAC exhibits more severe accuracy deterioration than the accuracy of the conventional operation. This highlights the effectiveness of the offset current compensation by accommodating a simple conductance offset mapping of VECOM.

\begin{figure}[t]
\vspace{-5mm}
\centerline{\includegraphics[width = .6\linewidth]{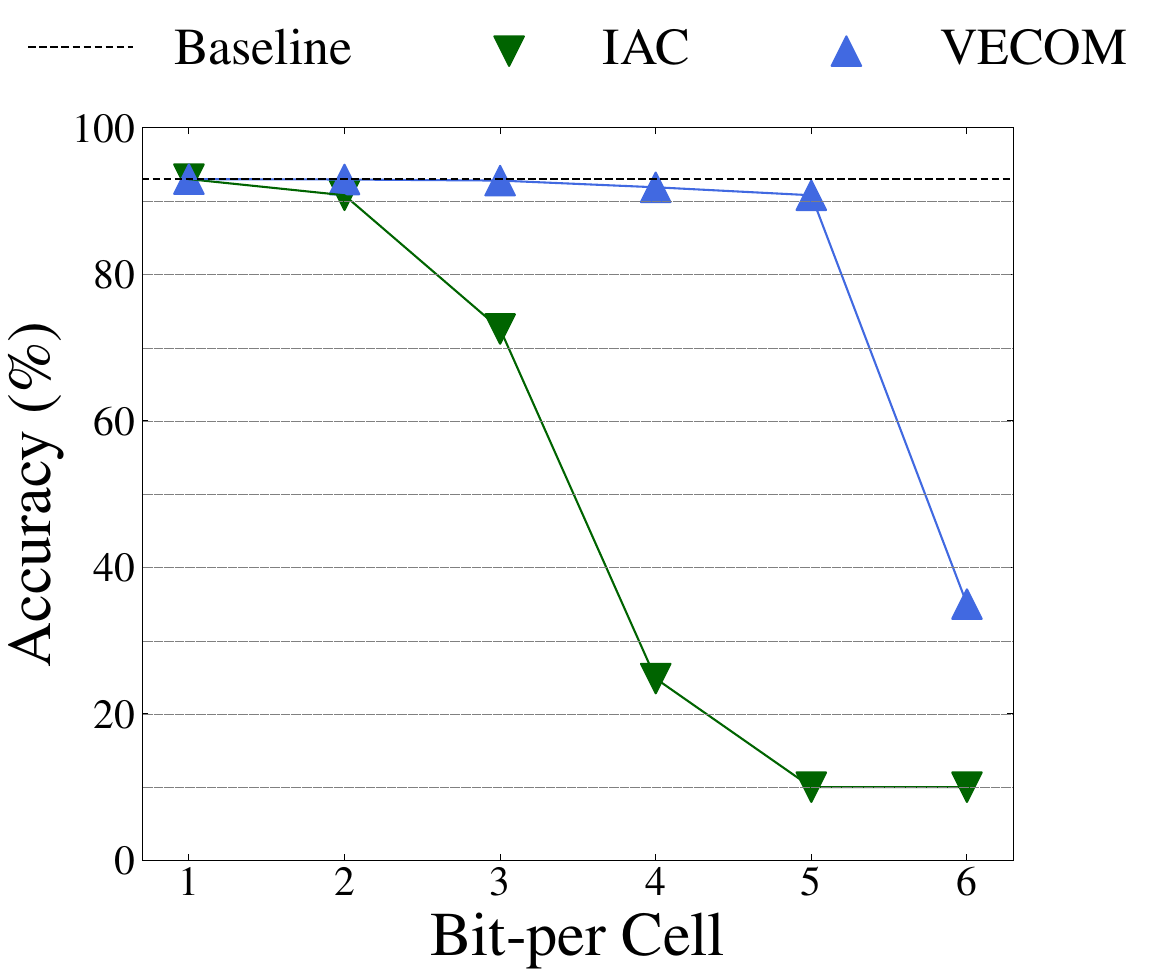}}

\caption{Scalability of VECOM on cell precision compared with IAC.}
\label{fig:cell_scale}

\end{figure}

Furthermore, to assess the scalability of VECOM in terms of bit-per-cell, we compare the accuracy of VECOM with the IAC technique in terms of bit-per-cell precision. In this experiment, the NAW is set to 128, the variation is set to 0.04, and the Bit-per-Cell is scaled from 1 to 6. The R-Ratio is set to 300, which represents a typical ReRAM cell value \cite{li2015variation}. The accuracy degradation is measured and compared, as shown in Fig. \ref{fig:cell_scale}. The experimental results reveal that the performance of IAC is only valid for SLC and the accuracy starts to decline when using 2 bits-per cell. However, VECOM shows less than 1\% difference from the baseline accuracy, even when extended to QLC with up to 4 Bit-per-Cell. This indicates that the current compensation can be successfully incorporated, even if future ReRAM processes employ cells with levels above MLC.

\begin{figure}[t]
\vspace{-5mm}
\centerline{\includegraphics[width = .8\linewidth]{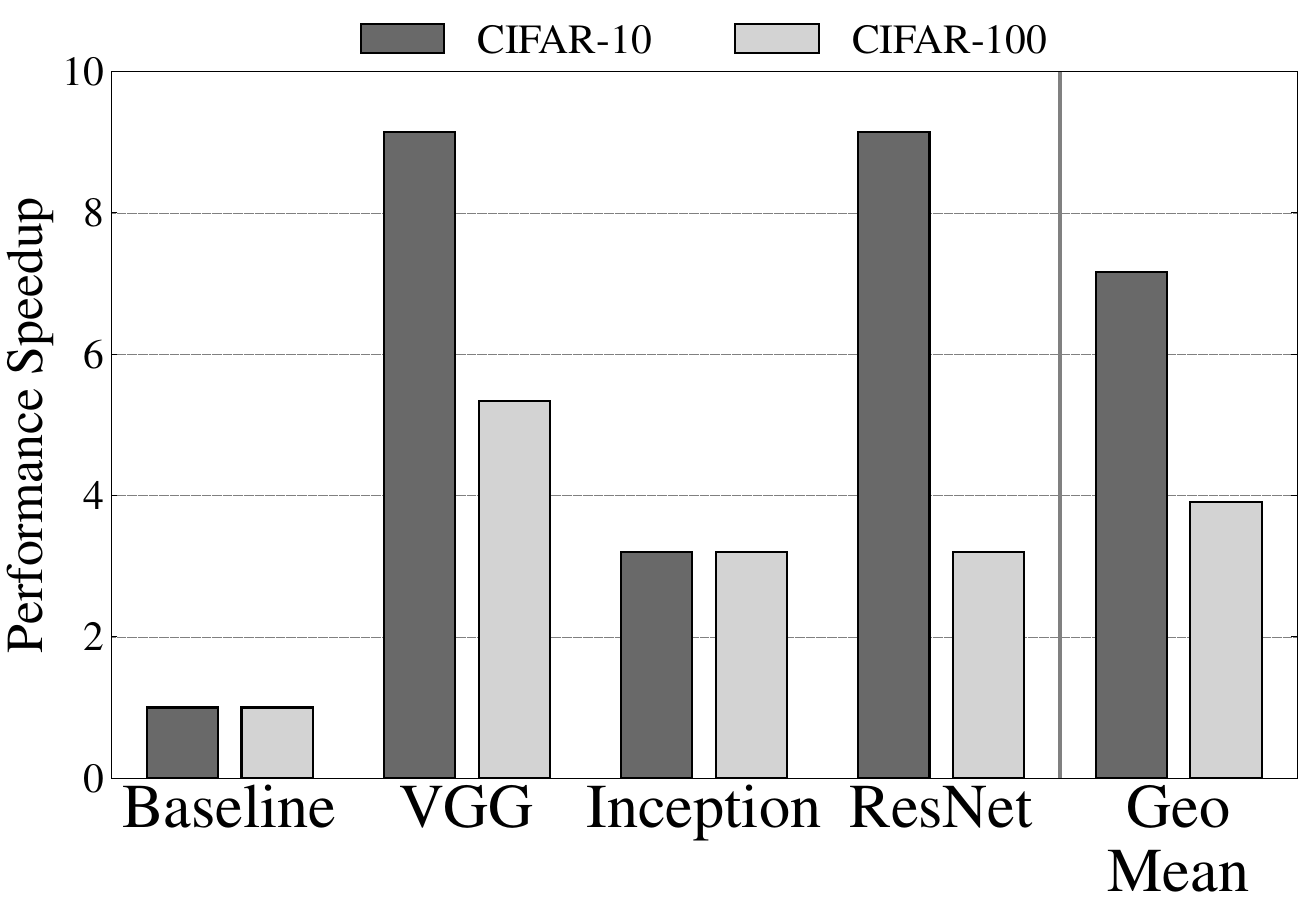}}

\caption{Performance speedup of VECOM on CIFAR-10 and CIFAR-100 datasets compared to the baseline.}
\label{fig:speedup}

\end{figure}

\subsubsection{Performance Improvement}\hfill\\
We evaluate the performance speedup of VECOM by comparing the number of cycles required to complete the crossbar array MAC operation with the maximum number of activated wordlines obtained in Sec. \ref{sec:variation_tolerance}. Since the conventional mapping results of Sec. \ref{sec:variation_tolerance} show a considerable accuracy degradation even with only 8 activated wordlines, we extrapolate them to ascertain the NAW at 1\% accuracy drop. For a fair comparison, we assume the use of a Successive Approximation Register (SAR) ADC and set the resolution of the ADC to $log_2(NAW) + 2$ bits. It is worth noting that the performance of the ADC's bit resolution is considered as the bottleneck in improving the performance of ReRAM-based DNN accelerators \cite{10.1145/3007787.3001140} hence the overall cycle time for the computation is dictated by the resolution of ADC\cite{10.1145/3307650.3322271}. Given these assumptions, we calculate the performance speedup of VECOM, as depicted in Fig. \ref{fig:speedup}. The estimations demonstrate that VECOM can achieve a speedup of up to 9.1 times (with an average of 7.2 times) on the CIFAR-10 dataset and up to 5.3 times (with an average of 3.9 times) on the CIFAR-100 dataset, compared to the baseline. VECOM amplifies the utilization of concurrent MAC results that the ADC can process by enabling more NAWs, which contributes to performance improvement.

\subsubsection{Energy Efficiency}
\begin{figure}[t]
\vspace{-5mm}
\centerline{\includegraphics[width = .8\linewidth]{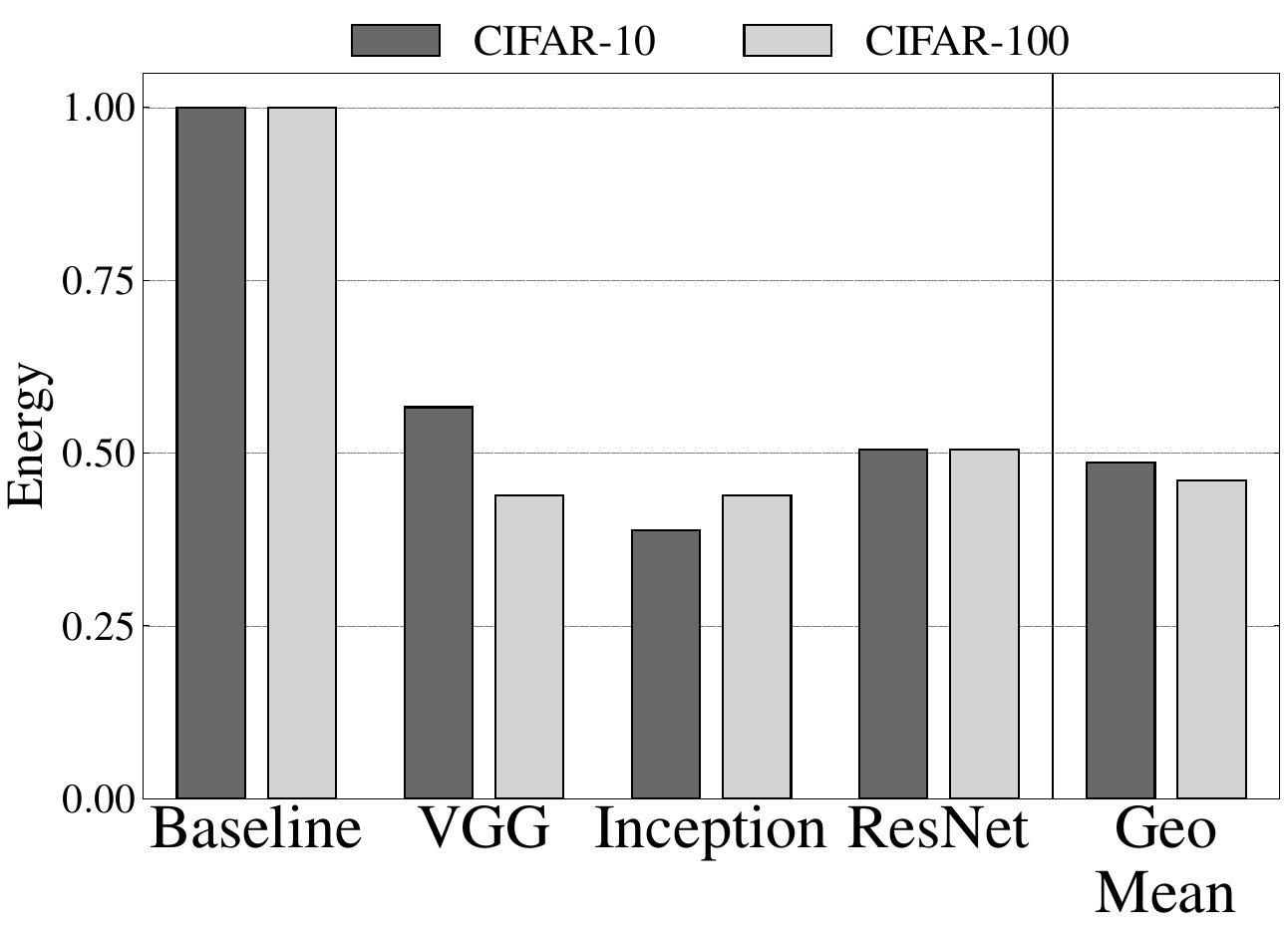}}
\vspace{-1mm}
\caption{Energy consumption of VECOM on CIFAR-10 and CIFAR-100 datasets compared to the baseline.}
\label{fig:energy}

\end{figure}

\hfill\\
VECOM's variation-resilient encoding strategy effectively maps conductance levels with significant variations to lower levels that exhibit less variation. This approach cuts down on the dynamic power consumption, proportional to the conductance value of the ReRAM cell, during MAC operations on the ReRAM crossbar array. Moreover, it enhances wordline-level parallelism, leading to improved performance and allowing the use of higher-resolution ADCs. To evaluate the energy consumption changes, we follow the power scaling equation described in \cite{5711005} based on the resolution of the SAR ADC and estimate dynamic energy consumption in the crossbar array. Fig. \ref{fig:energy} shows the results of comparing energy consumption using conventional mapping across various models and datasets. In these comparisons, we consider the energy expended when a single MAC operation in one array is completed as the baseline point.

Our empirical studies demonstrate that VECOM achieves an average reduction in operational energy consumption of over 50\% compared to the baseline. This reduction is achieved despite the increase in ADC power due to a higher ADC resolution. By exploiting a larger amount of accumulated current at once, the energy necessary to complete the operation decreases. Therefore, VECOM offers a clear advantage by lessening energy consumption while facilitating parallel operation.

The substantial energy savings achieved by VECOM further emphasize its potential for energy-efficient computing in ReRAM-based DNN accelerators. By mitigating the impact of variations and maximizing the utilization of accumulated current, VECOM introduces an efficient encoding scheme that contributes to the overall energy reduction, making it a promising technique for low-power and energy-constrained applications.

\section{Conclusion}\label{sec:conclusion}
This paper proposes VECOM, a variation-resilient encoding and offset compensation scheme for ReRAM-based DNN accelerators. Two techniques in VECOM are designed to address the reliability issues that often plague ReRAM devices in memory-intensive applications like DNNs. Our experimental results demonstrate that VECOM can significantly improve the throughput of ReRAM-based PIM architectures by up to 9.1 times without any software overhead. Additionally, VECOM improves energy efficiency and reliability against variation in ReRAM PIM operations. VECOM is designed to be lightweight and easily implementable, making it suitable for deployment in resource-constrained devices. Moreover, the increased sparsity resulting from VECOM's encoding method can be effectively combined with model compression techniques, further boosting performance gains. 
Another noteworthy aspect of VECOM is its stability even at low R-ratios and cell precision levels beyond multi-level cell (MLC). This makes it applicable to advanced ReRAM technologies that may emerge in future studies. We firmly believe that the proposed techniques in VECOM hold substantial practical implications for real-world scenarios involving ReRAM-based PIM architectures in memory-intensive applications like DNNs. By addressing reliability concerns through VECOM, we can achieve significant improvements in performance speedup, energy efficiency, accuracy, and reliability, thereby paving the way for the widespread adoption of ReRAM-based PIM architectures in memory-intensive applications.

\section{Acknowledgement}\label{sec:acknowledgement}
This work was supported in part by the Institute of Information \& Communications Technology Planning \& Evaluation (IITP) grant funded by the Korea Government (MSIT) under Grant 2022-0-00971 and 2021-0-00106; in part by the Next Generation Intelligent Semiconductor Development by the Ministry of Trade, Industry and Energy (MOTIE) under Grant 20011074; and in part by Samsung Electronics. The EDA Tools used in this work were supported by IDEC, Daejeon, South Korea.



\end{document}